\documentclass[aps,twocolumn,nofootinbib]{revtex4}
\usepackage[latin1]{inputenc}
\usepackage{epsfig}
\usepackage{amsfonts}
\usepackage{amssymb}
\usepackage{calrsfs}
\usepackage{amsmath}
\newcommand{\beq}{\begin{equation}}
\newcommand{\eeq}{\end{equation}}
\newcommand{\kB}{k_{\mbox{\tiny B}}}

\begin{document}

\title{Entropy production and heat capacity of systems under
time-dependent oscillating temperature}

\author{Carlos E. Fiore and Mário J. de Oliveira}
\affiliation{Universidade de São Paulo,
Instituto de Física,
Rua do Matão, 1371, 05508-090
São Paulo, SP, Brasil}

\begin{abstract}

Using the stochastic thermodynamics, we determine the entropy
production and the dynamic heat capacity of systems
subject to a sinusoidally time dependent temperature,
in which case the systems are permanently out of thermodynamic equilibrium
inducing a continuous generation of entropy. 
The systems evolve in time according to a Fokker-Planck 
or to a Fokker-Planck-Kramers equation. Solutions 
of these equations, for the case of harmonic forces,
are found exactly from which the heat flux, the production
of entropy and the dynamic heat capacity are obtained
as functions of the frequency of the temperature modulation.
These last two quantities are shown to be related to the
real an imaginary parts of the complex heat capacity. 

\end{abstract}

\maketitle

%--------------------------------------------------------------------
\section{Introduction}

The investigation of systems under time dependent fields of various types
is very common in experimental physics. Less common is the investigation
of systems under time dependent temperature. Nevertheless,
temperature oscillations is the basis of modulation calorimetry 
\cite{hohne2003,kraftmakher2004,filippov1966,sullivan1968,gobrecht1971,
birge1985,gill1993,schawe1995,jeong1997,schawe1997,hohne1997,simon1997,
jones1997,baur1998,claudy2000,garden2007a,garden2007c,garden2008},
which allows the experimental determination of the heat capacity.
The method consists in heating a sample by a periodical heating power 
with an angular frequency $\omega$ and measuring the temperature oscillations. 
This procedure induces a flow of heat from which the {\it dynamic}
heat capacity $C$ can be obtained as the ratio between the heat flux
$\Phi_{\rm q}$ and the time variation of the temperature, 
\beq
C=\frac{-\Phi_{\rm q}}{dT/dt}.
\label{0a}
\eeq

The heat flux as well as the heat capacity oscillate in time with
the same frequency $\omega$ of the temperature oscillations
but with a phase shift.
During a cycle the net heat flux vanishes but not the dynamic
heat capacity. Denoting by a bar the time average of a 
quantity, which is its integral over a cycle divided by the
period of the cycle, then $\overline{\Phi}_{\sf q }=0$ and
$\overline{C}$ is nonzero and shows a dispersion,
that is, a dependence on $\omega$. The conventional heat
capacity $C_0$, or static heat capacity, is obtained in the
limiting value of $\overline{C}$ when $\omega\to0$.

Under a time oscillating temperature, the system is permanently
out of equilibrium causing a continuous production of entropy
as well as a continuous flux of entropy. The entropy
$S$ of the system also varies in time, the time variation
being equal to the rate of entropy production $\Pi$ minus the 
entropy flux $\Phi$, 
\beq
\frac{dS}{dt} = \Pi - \Phi.
\label{21}
\eeq
According to the second law of thermodynamics,
the rate of entropy production is never negative, $\Pi\ge0$, 
but the flux of entropy $\Phi$, given by
\beq
\Phi = \frac{\Phi_{\sf q}}{T},
\label{0b}
\eeq
may have either sign. Although $\overline{\Phi}_{\sf q }=0$, this is not
the case of $\overline{\Phi}$. In fact, considering 
that the entropy $S$ is periodic, the left-hand side of (\ref{21})
vanishes in a cycle and the net flux becomes equal to
the entropy produced during a cycle, that is,
$\overline{\Phi}=\overline{\Pi}\geq0$.

Our main purpose here is the calculation of the entropy production and
the dynamic heat capacity for
systems subject to a temperature modulation of the type
\beq
T = T_0 + T_1\cos\omega t,
\eeq
where $T_1$ is the amplitude of modulation and $T_0$ is the
mean temperature. Our calculation is based on stochastic
thermodynamics of systems with continuous space of states 
\cite{tome2006,tome2010,broeck2010,spinney2012,zhang2012a,seifert2012,
santillan2013,luposchainsky2013,wu2014,tome2015,tome2015book}.
We restrict ourselves to the case of systems of particles
interacting through harmonic forces, in which case the 
evolution equation can be solved exactly.
From its solution, we determine the rate of entropy
production and dynamic heat capacity as a function of the
frequency $\omega$.  
We also show that the dynamic heat capacity and the
entropy production are related to the real and imaginary
parts of the complex heat capacity, respectively.

%--------------------------------------------------------------------
\section{Fokker-Planck equation}

\subsection{General formulation}

We consider a system of interacting particles that is described by
a probability distribution $P(x,t)$ of state $x$ at time $t$, where $x$
denotes the collection of particle positions $x_i$.
We assume that the time evolution of the probability distribution is
governed by the Fokker-Planck (FP) equation \cite{tome2006,tome2015book}
\beq
\frac{\partial P}{\partial t} = -\sum_i \frac{\partial J_i}{\partial x_i},
\label{54}
\eeq
where
\beq
J_i = \frac{1}{\alpha}(f_i P - \kB T \frac{\partial P}{\partial x_i}) , 
\eeq
and $f_i=-\partial V/\partial x_i$ is the force acting on particle $i$,
$V$ being the potential energy,
$\alpha$ is a constant and $\kB$ is the Boltzmann constant. 

The FP equation describes the contact
of the system with a heat reservoir at temperature $T$. 
Indeed, it is easily shown by replacement into the FP equation
that the Gibbs distribution
\beq
P_0 = \frac1Z e^{-V/\kB T}
\eeq
is the stationary solution when $T$ is kept constant
and, in fact, the equilibrium solution.

The time variation of the energy  $U=\langle V \rangle$
of the system can be obtained from the FP equation and is
\beq
\frac{dU}{dt} = - \Phi_{\rm q},
\eeq
where $\Phi_{\rm q}$ is the heat flux {\it from} the system
to outside and is expressed by \cite{tome2006},
\beq
\Phi_{\rm q} = \frac{1}{\alpha}\sum_i  (\langle f_i^2\rangle + \kB T \langle f_{ii}\rangle),
\label{55}
\eeq
where $f_{ii}=\partial f_i/\partial x_i$. 
Once the heat flux is known, the dynamic heat capacity is determined by
\beq
C = \frac{-\Phi_{\rm q}}{dT/dt},
\label{50a}
\eeq
if $T$ is time dependent.

From the FP equation we can also determine the time variation of
the entropy
\beq
S = -\kB \int P\ln P dx,
\eeq
which can be split in two terms,
\beq
\frac{dS}{dt} = \Pi - \Phi,
\eeq
where $\Pi$ is the rate of entropy production 
having the following form \cite{tome2006}
\beq
\Pi = \frac{\alpha}T  \sum_i \int \frac{J_i^2}{P} dx,
\eeq
and $\Phi$ is the entropy flux {\it from} the system to the 
environment 
\beq
\Phi 
= \frac{\Phi_{\rm q}}{T}.
\label{50b}
\eeq

%------------------------------------------------
\subsection{Harmonic forces}

When the forces are harmonic it is possible to exactly
solve the FP equation even for the case of a time
dependent temperature. Here, we consider a collection
of independent harmonic oscillators in which case it
suffices to treat just one oscillator. The
potential energy of the oscillator is
$V = k x^2/2$ which yields a force $f = - kx$
and the FP equations to be solved is
\beq
\frac{\partial P}{\partial t} = - \frac{\partial J}{\partial x},
\label{62}
\eeq
where
\beq
J = - \frac{k}{\alpha} x P - \frac{\kB T}{\alpha} \frac{\partial P}{\partial x}.
\eeq

The solution of the FP equation for a time dependent
temperature is a Gaussian distribution
\beq
P = \frac{1}{\zeta}\exp\{-\frac12 b x^2\},
\eeq
where the coefficients $b$ is time dependent.
That $P$ is a solution can be checked by replacing it
into the FP equation (\ref{62}).  Instead of seeking for the
coefficients $b$, we choose to find the 
averages $B=\langle x^2\rangle$. Once
$B$ is found we may get $b$, if necessary, from the relation $b=1/B$.

From the FP equation, we find the equation for $B$
\beq
\alpha\frac{d}{dt} B = - 2 k B + 2 \kB T.
\label{56}
\eeq
For $T$ depending on time like
\beq
T = T_0 + T_1\cos\omega t,
\eeq
the solution of equation (\ref{56}) is found to be
\beq 
B = \frac{\kB T_0}{k}
+ 2\kB T_1 \frac{2k\cos\omega t + \alpha\omega\sin\omega t}
{\alpha^2\omega^2+4k^2}.
\eeq

%------------------------------------------------
\subsection{Entropy production and heat capacity}

From equation (\ref{55}), it follows that the heat flux 
is determined by 
\beq
\Phi_{\rm q} =  \frac{k}\alpha (k B - \kB T ),
\eeq
or in an explicit form as
\beq
\Phi_{\rm q} = \kB T_1\omega k
\frac{2k\sin\omega t - \alpha\omega \cos\omega t}
{\alpha^2\omega^2 + 4k^2}.
\label{69}
\eeq
The entropy flux $\Phi$ and the dynamic heat capacity $C$ 
are determined from $\Phi_{\rm q}$ by the use of equations 
(\ref{50b}) and (\ref{50a}).

We proceed now to determine the time averages of $\Phi$  and $C$.
The time average of the heat flux vanishes
$\overline{\Phi}_{\rm q}=0$ as expected but not $\overline{\Phi}$
and $\overline{C}$. Carrying out the integration of $\Phi$ and $C$
over a cycle, and considering that $\overline{\Phi}=\overline{\Pi}$,
we find
\beq
\overline{\Pi} = \kB \lambda\frac{\alpha\omega^2 k}
{\alpha^2\omega^2+4 k^2},
\label{57}
\eeq
where
\beq
\lambda=\frac{T_0}{\sqrt{T_0^2-T_1^2}} - 1,
\label{43}
\eeq
and the dynamic heat capacity is found to be
\beq
\overline{C} = \kB \frac{2 k^2}{\alpha^2\omega^2+4 k^2}.
\label{58}
\eeq

%------------------------------------------------
\subsection{Harmonic oscillator}

The approach we have used above, by employing the FP equation (\ref{54})
or (\ref{62}), is appropriate do describe overdamped systems. 
In this approach the positions were taken into account 
but not the velocities. However, the oscillations of temperature
affect not only the positions but also the velocities of 
particles. The treatment of the response of the system 
concerning the velocities is carried out by setting up
the following FP equation that gives the evolution of the
probability distribution of velocites,
\beq
\frac{\partial P}{\partial t} = - \frac{\partial J}{\partial v},
\label{64}
\eeq
where
\beq
J = - \gamma v P - \frac{\gamma \kB T}{m}\frac{\partial P}{\partial v},
\eeq
which describes a free particle in contact with a reservoir at
a temperature $T$.

Equation (\ref{64}) is formally identical to equation (\ref{62})
and we may proceed in a similar way to determine
the entropy production and the heat capacities.
% x --> v,    k/alpha --> gamma,   1/alpha --> gamma/m
% ou   alpha --> m/gamma,   k --> m
The result for the heat flux is
\beq
\Phi_{\rm q} = \kB T_1\omega \gamma
\frac{2\gamma\sin\omega t - \omega \cos\omega t}{\omega^2 + 4\gamma^2}.
\label{69a}
\eeq
and the time average of the rate of entropy is
\beq
\overline{\Pi} = \kB \lambda\frac{\gamma\omega^2}{\omega^2+4\gamma^2},
\label{57a}
\eeq
where $\lambda$ is given by (\ref{43}),
and the dynamic heat capacity is
\beq
\overline{C} = \kB \frac{2\gamma^2}{\omega^2+4\gamma^2}.
\label{58a}
\eeq

To find the entropy production of a harmonic oscillator we should
add the entropy production concerning the positions,
given by (\ref{57}), with the entropy production concerning the velocities,
given by (\ref{57a}). The result is
\beq
\overline{\Pi} = \kB \lambda\frac{\gamma\omega^2 \kappa}
{\gamma^2\omega^2+4 \kappa^2} + \kB \lambda\frac{\gamma\omega^2}{\omega^2+4\gamma^2}.
\label{57b}
\eeq
Similarly, the dynamic heat capacity 
is the sum of (\ref{58}) and (\ref{58a}),
\beq
\overline{C} = \kB \frac{2 \kappa^2}{\gamma^2\omega^2+4 \kappa^2}
+ \kB \frac{2\gamma^2}{\omega^2+4\gamma^2}.
\label{58b}
\eeq
The quantities $\alpha$ and $\gamma$ are related to
$\alpha=m \gamma$, and $k$ is related to $\kappa$ by $k=m\kappa$.

\begin{figure*}
\epsfig{file=capov.eps,width=8.7cm}
\hfill
\epsfig{file=imaov.eps,width=8.7cm}
\caption{Real (a) and imaginary (b) parts of the complex
heat capacity (\ref{63}) for the overdamped case 
as a function of frequency for the
following values of $\kappa/\gamma^2$: 0 (dotet line), 0.1, 0.2, 0.5, 1, and 2
(from left to right).}
\label{overdamp}
\end{figure*}

\begin{figure}
\epsfig{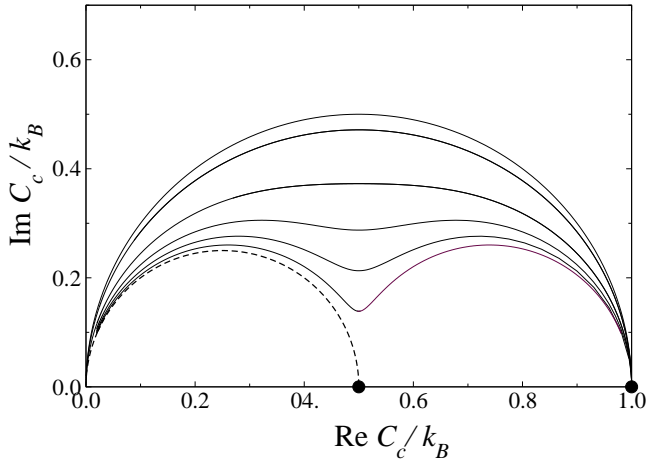}
\caption{Imaginary versus real part of the complex
heat capacity (\ref{63}) for the overdamped case for the
following values of $\kappa/\gamma^2$: 0 (dottet line), 0.02, 0.05, 0.1, 0.2, 0.5, 1
(from bottom to top). The thermodynamic equilibrium, $\omega=0$, is indicated by a 
full circle.}
\label{ixrover}
\end{figure}

%------------------------------------------------
\subsection{Complex heat capacity}

The dispersion of the dynamic heat capacity on frequencies,
induced by a time varying temperature, 
has an analogy with the dispersion of susceptibility on
frequencies induced by a time varying field. In this last
case, the response to the field oscillation is described by a complex
susceptibility. Analogously, it is also possible
to define a complex heat capacity to conveniently describe
the response to temperature oscillations.
In fact, the complex heat
capacity has been the subject of investigation in relation to
temperature modulation \cite{gobrecht1971,birge1985,
gill1993,schawe1995,jeong1997,schawe1997,hohne1997,simon1997,
jones1997,baur1998,claudy2000,garden2007a,garden2007c,garden2008}

Suppose that we replace $T$ in equation (\ref{56}) by
\beq
T_c = T_0 + T_1 e^{-i\omega t}.
\eeq
Then, instead of equation (\ref{69}) and (\ref{69a}), 
we would get the following expression for the heat flux of
the harmonic oscillator, 
\beq
\Phi_{\rm q}^c = \kB T_1 
\left(\frac{i\kappa \omega}{2 \kappa - i\omega\gamma}
+ \frac{i \gamma \omega}{2\gamma - i\omega}\right)
e^{-i\omega t},
\eeq
By analogy with (\ref{50a}), a complex heat capacity $C_c$ can be defined by
\beq
C_c = \frac{-\Phi_{\rm q}^c}{dT_c/dt},
\eeq
from which we find
\beq
C_c = \kB \left(\frac{\kappa}{2 \kappa - i\omega\gamma}
+ \frac{\gamma}{2\gamma - i\omega}\right)
\label{63}
\eeq
which is time independent.
Comparing with expressions (\ref{57b}) and (\ref{58b}), we see that
\beq
\overline{C} = \Re(C_c)
\qquad\qquad
\overline{\Pi} = \frac1{\lambda\omega} \Im(C_c)
\eeq
These results show that the real part of the complex
heat capacity is identified with the dynamic heat capacity
and the imaginary part is proportional to the rate of
entropy production.

The real and imaginary parts of the complex heat capacity $C_c$
are shown in Fig. \ref{overdamp} as functions of the frequency
$\omega$ for several values of $\kappa$. The real part, which is the
dynamic heat capacity $\overline{C}$, becomes the static heat capacity
when $\omega\to0$, which is $C_0=\kB/2$ if $\kappa=0$ and $C_0=\kB$
if $\kappa\neq0$. In the opposite limit, $\omega\to\infty$, it
vanishes as $1/\omega^2$. The imaginary part vanishes when $\omega\to0$
and so does the rate of entropy production $\overline{\Pi}$. In the
limit $\omega\to\infty$, the imaginary part vanishes as $1/\omega$
but the rate of entropy production reaches a finite value, which is
$\overline{\Pi}=\kB \lambda(\gamma +\kappa/\gamma)$.
In Fig. \ref{ixrover} we have plotted $\Im(C_c)$ versus $\Re(C_c)$
and we sees that the curves are symmetric.

%-------------------------------------------------------------------
\section{Fokker-Planck-Kramers equation}

\subsection{General formulation}

We consider again a system consisting of several interacting particles 
in contact with a temperature reservoir at temperature $T$,
with which it exchanges heat. The time
evolution of the probability  distribution $P(x,v,t)$,
where $x$ denotes the collection of the positions $x_i$
and $v$ the collection of velocities $v_i$ of the particle,
is governed by the Fokker-Planck-Kramers (FPK) equation
\cite{tome2010,tome2015,tome2015book}
\beq
\frac{\partial P}{\partial t} = - \sum_i\left(
v_i\frac{\partial P}{\partial x_i}
+ \frac1{m}f_i \frac{\partial P}{\partial v_i}
+ \frac{\partial J_i}{\partial v_i}\right),
\label{3}
\eeq
where
\beq
J_i = - \gamma v_i P - \frac{\gamma \kB T}{m}
\frac{\partial P}{\partial v_i}.
\label{3a}
\eeq
Here, $m$ is the mass of each particle, $\gamma$ is the dissipation constant,
and $f_i$ the force acting on the particle $i$,
given by $f_i=-\partial V/\partial x_i$.

If the temperature $T$ is kept constant, 
then for large times the probability distribution
approaches the Gibbs equilibrium distribution,
\beq
P^{e}(x,v) = \frac{1}Z e^{-E/\kB T},
\label{9}
\eeq
where $E=mv^2/2 + V$ is the energy of the system.
This result shows that the FPK equation (\ref{3}) indeed 
describes the contact of a system with a heat reservoir at a
temperature $T$.

The time variation of the energy $U=\langle E\rangle$ 
is obtained from the FPK equation and is 
\beq
\frac{dU}{dt} = - \Phi_{\rm q},
\label{10}
\eeq
where the heat flux $\Phi_{\rm q}$ {\it from} the
system to outside is expressed as \cite{tome2010,tome2015}
\beq
\Phi_{\rm q} = \sum_i(\gamma m \langle v_i^2\rangle - \gamma \kB T),
\label{11}
\eeq
where the first and second terms are understood as the
heating power and the power of heat losses, respectively,
with $\gamma\kB$ being the heat transfer coefficient \cite{kraftmakher2004}. 

The entropy $S$ of the system is determined from the
Gibbs expression
\beq
S = - \kB \int P \ln P dxdv.
\label{17}
\eeq
Using the FPK equation, one finds that
its time derivative can be split into two terms,
\beq
\frac{dS}{dt} =\Pi - \Phi,
\eeq
where the first is the rate of entropy production 
which can be written as \cite{tome2010,tome2015}
\beq
\Pi = \frac{m}{\gamma T} \sum_i \int \frac{J_i^2}{P}dxdv,
\eeq
and the second is the flux of entropy which can be written
in the following form
\beq
\Phi = \frac{\Phi_{\rm q}}{T},
\label{14}
\eeq
where $\Phi_{\rm q}$ is the heat flux given by (\ref{11}).
If $T$ is time dependent then the dynamic heat capacity
is obtained by
\beq
C = \frac{-\Phi_{\rm q}}{dT/dt}.
\label{14a}
\eeq

\begin{figure*}
\epsfig{file=cap.eps,width=8.7cm}
\hfill
\epsfig{file=ima.eps,width=8.7cm}
\caption{Real (a) and imaginary (b) parts of the complex
heat capacity (\ref{67}) as a function of frequency for the
following values of $\kappa/\gamma^2$: 0 (dottet line), 0.1, 0.2, 0.5, 1, 2, and 5
(from left to right).
}
\label{complex}
\end{figure*}

\begin{figure}
\epsfig{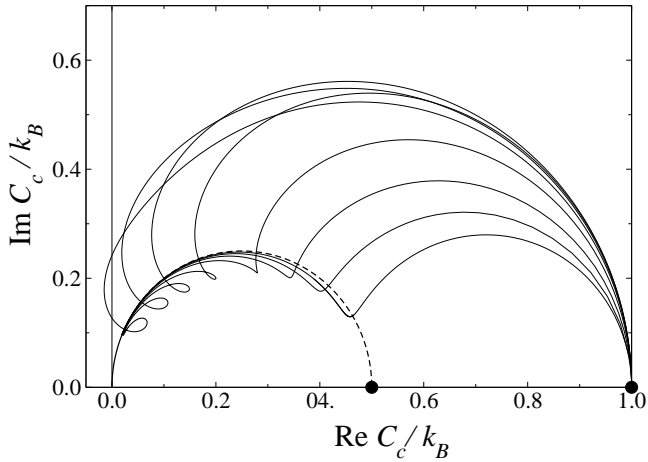}
\caption{Imaginary versus real part of the complex
heat capacity (\ref{67}) for the
following values of $\kappa/\gamma^2$: 0 (dottet line), 
0.02, 0.05, 0.1, 0.2, 0.5, 1, 2, and 5 (from right to left).
The thermodynamic equilibrium, $\omega=0$, is indicated by a 
full circle.}
\label{ixr}
\end{figure}

%------------------------------------------------
\subsection{Harmonic oscillator}

We consider here the case of just one harmonic oscillator.
When the temperature or the external force is time dependent,
the probability distribution (\ref{9}) is no longer
the solution of the Fokker-Planck equation for long times,
and we should seek for a solution. 
When the force is harmonic, which we write as $f=-m\kappa x$,
the FPK equation can be solved exactly.
The solution is a Gaussian distribution in $x$ and $v$
of the type
\beq
P(x,v) = \frac{1}{\zeta} \exp\{- \frac12(a v^2 + b x^2 + 2cxv) \},
\label{26}
\eeq
where the parameters $a$, $b$, and $c$ depend on time.
That this Gaussian distribution is a solution can be
checked by replacing it into the FPK equation.
The solution is reduced to the determination of the time
dependence of the parameters.

From the Gaussian distribution (\ref{26}) we see that
the parameters $a$, $b$, and $c$ are related to
the averages $A=\langle v^2\rangle$,
$B=\langle x^2\rangle$ and $C=\langle xv\rangle$ as follows
\beq
a = \frac{B}{AB-C^2}, \qquad
b = \frac{A}{AB-C^2}, \qquad
c = \frac{C}{AB-C^2}.
\eeq
The method we use here rests on setting up equations
for $A$, $B$, and $C$, from whose solutions we can find
the coefficients $a$, $b$, and $c$ of the Gaussian distribution
as functions of temperature, if needed.

From the FPK equations the following set of equations 
are found for $A$, $B$, and $C$
\beq
\frac{dA}{dt} = -2\kappa C - 2\gamma A + \frac{2\gamma \kB T}{m},
\label{19a}
\eeq
\beq
\frac{dB}{dt} = 2 C,
\label{19b}
\eeq
\beq
\frac{dC}{dt} = A - \kappa B - \gamma C.
\label{19c}
\eeq

Equations (\ref{19a}), (\ref{19b}), and (\ref{19c})
are coupled linear differential equations whose solution
can also be found for a temperature modulation of the type
\begin{equation}
T = T_0 + T_1\cos\omega t.
\label{8}
\end{equation}
The solution of the set of equations (\ref{19a}), (\ref{19b}),
and (\ref{19c}) gives the following result for $A$,
\beq
A = \frac{\kB T_0}{m} + \frac{\kB T_1}{m} 
(A_1 \cos\omega t + A_2 \sin\omega t),
\label{22}
\eeq
\beq
A_1 = \frac{4\gamma^2(\omega^4 - 3\kappa\omega^2 + 4\kappa^2 + \gamma^2\omega^2)}
{(\omega^2+\gamma^2)[(\omega^2-4\kappa)^2 + 4\gamma^2\omega^2]},
\eeq
\beq
A_2 = \frac{2\gamma\omega(\omega^4 - 6\kappa\omega^2 + 8\kappa^2 + \gamma^2\omega^2)}
{(\omega^2+\gamma^2)[(\omega^2-4\kappa)^2 + 4\gamma^2\omega^2]}.
\eeq

%--------------------------------------------------------------------
\subsection{Entropy production and heat capacity}

Using the result (\ref{22}) for $A$, we can write the heat flux
\beq
\Phi_{\rm q} = \gamma (m A - \kB T),
\eeq
in the explicit form 
\beq
\Phi_{\rm q} = \kB T_1 \gamma [(A_1 - 1)\cos\omega t + A_2 \sin\omega t].
\label{61}
\eeq
The entropy flux $\Phi$ and the dynamic heat capacity $C$ are obtained from 
this expression for $\Phi_{\rm q}$ and by the use of equations 
(\ref{14}) and (\ref{14a}). To get the time averages of
$\Phi$ and $C$ we should integrate them over one cycle.
Carrying out the integration, and taking into account that
$\overline{\Phi}=\overline{\Pi}$, we find
\beq
\overline{\Pi} = \kB \lambda \gamma (1-A_1),
\label{65a}
\eeq
or in a explicit form
\beq
\overline{\Pi} = \kB \lambda \frac{\gamma \omega^2
(\omega^4 - 8\kappa\omega^2 + 16\kappa^2 + 4\kappa\gamma^2 + \gamma^2\omega^2)}
{(\omega^2+\gamma^2)[(\omega^2-4\kappa)^2+4\gamma^2\omega^2]},
\label{70a}
\eeq
where $\lambda$ is given by equation (\ref{43}), and
\beq
\overline{C} = \kB \frac{\gamma}{\omega} A_2,
\label{65b}
\eeq
or in a explicit form
\beq
\overline{C} = \kB  
\frac{2\gamma^2 (\omega^4 - 6\kappa\omega^2  + 8\kappa^2 + \gamma^2\omega^2 )}
{(\omega^2+\gamma^2)[(\omega^2-4\kappa)^2+4\gamma^2\omega^2]}.
\label{70b}
\eeq

The results above were obtained for the case of a harmonic
oscillator. It is possible to find the results for a free
particle by formally setting $\kappa=0$. Using this procedure
we recover the results (\ref{57a}) and (\ref{58a})
for a free particle.

%------------------------------------------------
\subsection{Complex heat capacity}

Again we may set up the complex heat capacity. 
If equations (\ref{19a}), (\ref{19b}), and (\ref{19c})
are solved by replacing the temperature $T$ by
\beq
T_c = T_0 + T_1 e^{-i\omega t}.
\eeq
then instead of expression (\ref{61}) we would get
\beq
\Phi_{\rm q}^c = \kB T_1 \gamma (A_1 -1 + i A_2) e^{-i\omega t}.
\label{61a}
\eeq
and the following complex heat capacity 
\beq
C_c = \kB \frac{\gamma}{i\omega} (A_1 -1 + i A_2),
\label{67}
\eeq
which is time independent. The real and imaginary parts
of $C_c$ are 
\beq
\Re(C_c) = \kB \frac{\gamma}{\omega}A_2,
\qquad
\Im(C_c) = \kB \frac{\gamma}{\omega} (1 - A_1),
\eeq
and using relations (\ref{65a}) and (\ref{65b}) we find
\beq
\Re(C_c) = \overline{C},
\qquad\qquad
\Im(C_c) = \frac{1}{\lambda\omega}\overline{\Pi}.
\label{60}
\eeq
Again, these results show that the real part of the complex
heat capacity is the dynamic heat capacity
and the imaginary part is proportional to the rate of
entropy production.

The real and imaginary parts of the complex heat capacity $C_c$
are shown in Fig. \ref{complex} as functions of the frequency
$\omega$ for several values of $\kappa$. The real part, which is the
dynamic heat capacity $\overline{C}$, becomes the static heat capacity
when $\omega\to0$, which is $C_0=\kB/2$ if $\kappa=0$ and $C_0=\kB$
if $\kappa\neq0$. In the opposite limit, $\omega\to\infty$, it
vanishes as $1/\omega^2$. The imaginary part vanishes when $\omega\to0$
and so does the rate of entropy production $\overline{\Pi}$. In the
limit $\omega\to\infty$, the imaginary part vanishes as $1/\omega$
but the rate of entropy production reaches a finite value, which is
$\overline{\Pi}=\kB \lambda\gamma$.
In Fig. \ref{ixr} we have plotted $\Im(C_c)$ versus $\Re(C_c)$.

%------------------------------------------------
\subsection{Dynamic heat capacity}

During a small interval of time $\Delta t$, the heat introduced
equals $-\Phi_{\rm q}\Delta t$, which divided by the increment
$\Delta T$ in temperature gives $\Phi_{\rm q}\Delta t/\Delta T$. 
The heat capacity is obtained by taking the limit 
$\Delta t\to0$, 
\beq
C= -\frac{\Phi_{\rm q}}{dT/dt},
\label{23}
\eeq
which is the expression of the dynamic heat capacity that 
we have used. In the absence of external work, which is the 
case of the present analysis,  
$-\Phi_{\rm q}=dU/dt$ and the heat capacity is related to
the energy by $C=(dU/dt)/(dT/dt)$.

The dynamic heat capacity does not share with the static
heat capacity $C_0$ the property $C_0\ge0$. Generically, the heat flux
is not in phase with the variation of temperature.
A flux of heat to outside could happen while 
the temperature is increasing, or a flux toward the system
could happen while the temperature is decreasing. In both cases
the dynamic heat capacity has a negative sign. 
This peculiar but not illegitimate behavior is shown
in Fig. \ref{complex}a for a small interval of frequencies
for one of the curves. Notice, on the other hand,
that the rate of entropy production is always nonnegative
as illustrated in Fig. \ref{complex}b.

%--------------------------------------------------------------------
\section{Conclusion}

We have determined the entropy production and
the dynamic heat capacity of systems under time varying
temperature by the use of stochastic thermodynamics. 
The systems that we have analyzed evolves in time
according to the Fokker-Planck, for the overdamped case,
or to the Fokker-Planck-Kramers
equation. Exact solutions were possible to find for the
cases of harmonic forces and temperature modulation 
of the sinusoidal type. The heat flux also varies
sinusoidally but with a phase shift with respect to
temperature. From the heat flux, the rate of entropy production
$\overline{\Pi}$ and the dynamic heat capacity $\overline{C}$
could be determined
as functions of the frequency $\omega$ of the temperature
modulation. In the limit of small frequencies,
$\overline{C}$ approaches the equilibrium heat capacity,
which is nonnegative, and vanishes for large frequencies.
The dynamic heat capacity may not be a monotonic decreasing
function of $\omega$ and might even be negative.
The rate of entropy production is always nonnegative,
vanishing for zero frequency, when the system is
in equilibrium. For large values of $\omega$ it approaches
a nonzero value. Finally, $\overline{C}$ and $\overline{\Pi}$
were shown to be related to the real an imaginary parts
of the complex heat capacity. 

The calculation of the production of entropy and heat capacity
that we made above can be extended to a harmonic solid whose
potential energy is
\beq
V = \frac12 \sum_{ij} K_{ij}x_ix_j,
\eeq
where $K_{ij}$ are the elements of a symmetric matrix with
nonnegative eigenvalues $m\kappa_\ell$. The harmonic solid
can be understood as a collection of independent harmonic
oscillators with distinct frequencies that are $\sqrt{\kappa_\ell}$.
Therefore, we expect the total entropy production and the
total heat capacity to be the sum of the entropy productions and
the heat capacities of the individual oscillator, each one
having a spring constant equal to $m\kappa_\ell$,
\beq
\overline{\Pi}_{\rm HS} = \sum_{\ell} \overline{\Pi}(\kappa_\ell),
\qquad\qquad
\overline{C}_{\rm HS} = \sum_{\ell} \overline{C}(\kappa_\ell),
\eeq
where $\overline{\Pi}(\kappa)$ and $\overline{C}(\kappa)$
are given by the expressions (\ref{70a}) and (\ref{70b}), 
or in the overdamped case by the expressions (\ref{70a}) and (\ref{70b}),
respectively.

%--------------------------------------------------------------------

\end{document}